# Substitution effects of Ag/Cu and Al/Cu on Y123 samples: Hall anomaly and transport properties

M. Nazarzadehmoafi[1], V. Daadmehr[*1], A. T. Rezakhani[2], S. Falahati[1], F. Saeb[1], and S. Barekat Rezaei[1]

1. Magnet & Superconducting Research Lab, Department of physics, Alzahra University, Tehran 19938, Iran
2. Department of Physics, Sharif University of Technology, Tehran, Iran

**Abstract**

The Hall effect and its associated anomalies offer an important means for describing superconducting property in cuprate compounds. Here, we investigate the electron- and hole-doping effects (implemented through Al- and Ag-doping, respectively) on the Hall effect and transport properties of Y123 superconducting materials ($YBa_2Cu_{3-x}X_xO_{7-\delta}$; X= Ag, Al). The Hall anomaly, in both normal and vortex states, generally depends on multiple factors, such as temperature, magnetic field, and amount/type of doping. Specifically, for our samples, we observed that hole doping first enhances the critical temperature $T_C$, whereas its excess beyond a threshold value (here x=0.15) decreases $T_C$; electron doping, however, decreases $T_C$. We also show how doping changes the critical current density $J_C$ at temperature 77K with magnetic field 9kOe. More importantly, all of our samples exhibited one Hall sign reversal. Interestingly, behavior of the Hall coefficient vs. temperature reveals that the Hall dip vs. magnetic field for Ag – doped samples shows a behavior opposite to that for Al-doped samples.

## 1. Introduction

Transport measurements using the Hall effect are generally used to obtain physical properties, such as type and density of charge carriers, in normal-state superconductors. In high-$T_C$ superconductors (HTSCs), one of the striking and baffling properties is the linear behavior of the inverse Hall coefficient ($R_H$) versus temperature, $R_H^{-1} \propto T$, which is rare in metals [1], even in metals with complex Fermi surface, such as Cu, Ag, W, Mg, and Ca [2]. In HTSCs, this anomaly can be explained by considering the existence of two or more charge carriers, different temperature dependence of mobility of each band, or the thermal expansion (which produces a redistribution of the carriers among the bands). This peculiar temperature dependence is attributed to the existence of two relaxation times. Several theories have been suggested to explain the existence of these relaxation times; e.g., the Anderson model, which attributes the relaxation times to two quasiparticles (spinon and holon) [3], charge conjugation symmetry [4], skew scattering [5], and different T-dependence of the relaxation time on distinctive parts of the Fermi surface [6].

A puzzling phenomenon in vortex state of superconductors is sign reversal [7], including single, double [8-12], and triple reversals observed in the Hall effects in the superconducting state in most HTSCs and some conventional superconductors [13,14]. In normal state, the topology of the Fermi surface determines the Hall sign, whereas this sign is determined by the vortex motion in the superconducting state [13]. The origin of this anomaly is not yet known, although various models have offered partial explanations [15]. However, this anomaly contradicts conventional theories of flux motion, such as the Bardeen-Stephen [16] and the Nozieres-Vinen [17] models, which predict that the Hall effect stems from quasi-normal core and therefore should have the sign as in the normal state [18]. To elucidate this sign reversal, several theories, such as the time-dependent Ginsburg-Landau theory (TDGL) have been proposed [19, 20]. In the TDGL theory, the vortex Hall conductivity $\sigma^V_{xy}$ − originating from vortex mechanism contribution − plays an important role in determining the Hall sign at low fields. In fact, it implies that if $\sigma^V_{xy}$ has a sign opposite to the normal-state Hall effect, the sign reversal will occur [7]. The importance of the electronic structure in $\sigma^V_{xy}$ has already been argued in the literature [21-23]. Moreover, in a comprehensive survey [7], it has been shown that this anomaly is prevalent in HTSCs, and is determined by doping, magnetic field, and temperature.

There also exist theories emphasizing the role of backflow current due to pinning in the Hall sign-reversal phenomenon [24-26]. There, thermal fluctuations and forces acting on vortices (in the vortex state) − such as Lorentz, drag, Magnus, and pinning forces, and also vortex-vortex interaction − and the competition among such forces in the emergence of the sign reversal have been studied.

In this work, we demonstrate the Hall effect in fabricated superconducting samples with Ag-hole doping and Al-electron doping. Besides measuring various parameters, such as $T_C$ and $J_C$, and how temperature, magnetic field, and dopant (type and amount) affect such parameters, we investigate the doping effect in the emergence of the Hall sign reversal in vortex state. This study may shed some light on how pinning due to different doping contributes to the anomalous sign reversal phenomenon.

2. **Experiment**

In order to investigate the effects of X/Cu substitution (X=Ag, Al), we prepared single-phase polycrystalline $YBa_2Cu_{3-x}X_xO_{7-\delta}$ bulk samples with x=0, 0.1, 0.15, 0.2, and 0.3 stoichiometry for the Ag hole-doping amount, and x=0, 0.01, 0.02, 0.03, and 0.045 stoichiometry for the Al electron-doping amount by the sol-gel method. For precursor solutions, stochiometric amount of high purity (99.9%) Y, Ba, Cu, X (X=Ag, Al) nitrate in molar ratio |1: 2: 3-x: x| were dissolved in de-ionized water with 0.5M concentration. Citric acid as complexing agent and Ethylenediamin as neutralizer were used. The equivalent amount of metal nitrate and citric acid were also applied. All samples were prepared in pH=7 in light of the fact that we had already reported that the best superconducting properties could be obtained in this pH [27]. The solution was heated at 80˚C to make a viscous violet gel. The gel was heated in furnace at 520˚C for 2 hours. The powder calcined at 900˚C for 22 hours and was cooled down in air at furnace rate. The calcined compounds were grounded slightly and pressed into pellets of 1.3cm diameter at $10T/cm^2$ pressure. The pellets were sintered for 19 hours at 930˚C, cooled down rapidly to 630˚C for tetragonal to orthorhombic transition [8], and kept for 2 hours, and then annealed at the rate of 1˚C/min to room temperature. The sinter process took place in oxygen atmosphere. The structural phase analysis of the specimens was determined by powder X-ray diffraction patterns (XRD, Philips® Cu kα radiation). The Rietveld analysis of the XRD patterns was carried out by using the Material Analysis Using Diffraction (MAUD) program [28]. The microstructure was observed by scanning electron microscopy (SEM, Philips® XL30). The electrical resistivity was measured using the standard four-probe technique from 77-300K, and the I-V analysis was performed at 77K at the applied field of 9kOe. Temperature was monitored by a Pt-100 resistor. The ambient temperature in the range of 77-300K was provided by using liquid nitrogen. The Hall voltage measurements were executed with mA current, which was generated by Lake Shore®-120 current source at the applied magnetic field of H=2.52, 4.61, and 6.27kOe perpendicular to the transport current. Six-probe copper wires, attached by silver paste to the sample, were used to establish electrical contacts in order to cancel small offsets due to the mismatches of the Hall probe. The Hall coefficient was obtained from the transverse resistance by subtracting the positive and negative magnetic field data at a constant temperature.

3. Results and discussion

The structural phase analysis of the specimens was determined by powder X-ray diffraction patterns. Figure 1a and 1b show the XRD patterns of the Ag-doped samples with x=0, 0.1, 0.15, 0.2, and 0.3, and also the Al-doped samples with x=0, 0.01, 0.02, 0.03, and 0.045. It is seen that the XRD patterns of the Ag-doped samples (except for x=0.3) show a single orthorhombic phase without any impurity phases. It is thus evident that the Ag/Cu substitution has taken place completely and no impurity phases was established during this process. Three small peaks of Ag in the XRD pattern of the sample with x=0.3 indicate that the excess Ag doping does not enter the Y-123 structure, thus remains with grains and plays the role of intergrain weak links. The XRD patterns of the Al-doped samples with x=0, 0.01, 0.02 show a single orthorhombic phase, while the samples with x=0.03, 0.045 contain small number of slight $Y_2BaCuO_5$ (Y211) impurity peaks.

The microstructure was observed by SEM. In the SEM images, the improvement on grain structure is observed: the YBCO grains are strongly linked, and the grain size and homogeneity of the samples increase with Ag doping (Fig. 2). Ag modifies the intergrain connection from weak links to strong links, and is expected to increase the critical current density (see Ref. [29] for a similar effect of Ag doping on $J_C$). As indicated in Fig. 3, the SEM analysis of the $YBa_2Cu_{3-x}Al_xO_{7-\delta}$ samples shows that the homogeneity of the samples decreases, and porosity and grain size increase with Al doping (see also Ref. [28]).

The Rietveld analysis of the XRD patterns was carried out by using the MAUD program. The results of the XRD patterns of $YBa_2Cu_{3-x}Ag_xO_{7-\delta}$ are presented in Table 1. All samples are orthorhombic with Pmmm space group. The occupancy parameters of the XRD data refinement show that Ag occupies the Cu (1) site. Ag substitution for Cu (1) does not change the unit cell parameters significantly.

The results of the XRD refinement for $YBa_2Cu_{3-x}Al_xO_{7-\delta}$ are presented in Table 2. The orthorombicity parameter decreases with Al doping from 0.84 in an undoped sample to 0.59 in an Al-doped YBCO sample with x=0.045. Table 2 shows that, similarly to the case of Ag doping, Al occupies the Cu (1) site. Figure 4a and 4b depict the temperature dependence of the electrical resistivity in the Ag- and Al-doped samples. The normal-state resistance of the Ag-doped samples indicates a metallic behavior. The transition temperature ($T_C$) of the Ag-doped samples increases with Ag doping up to x=0.15 and decreases for excess Ag doping in samples with x=0.2, 0.3. In addition, all $YBa_2Cu_{3-x}Ag_xO_{7-\delta}$ samples present sharp transition to the superconducting state − the transition width ($\Delta T_C$) remains constant and shows that Ag does not perform any impurity phases (also shown by the XRD refinement).

We thus conclude that x=0.15 is the optimum value of the Ag-doping concentration, which shows the maximum $T_C$ in the Ag-doped YBCO samples (see also Ref. [31], in which a $T_C$ enhancement due to Ag doping has been reported). The R-T measurements of the $YBa_2Cu_{3-x}Al_xO_{7-\delta}$ samples in the normal state indicate metallic behavior for the samples (in good agreement with Ref.

[32]). Figure 4b shows that $T_C^{(zero)}$ first changes slightly by increasing the Al-doping concentration (up to x=0.02), while after that it follows a sharp variation. We suggest that placement of Al at the Cu (1) site localizes holes at the Cu-O chains − in agreement with Refs. [33,34]. Note that the decrease of $T_C$ with increasing Al doping is also in agreement with Refs. [35,36].

For studying of the doping dependence of the critical current density $J_C$, we used I-V measurements. Figure 5a and 5b show the E-J curves of the $YBa_2Cu_{3-x}X_xO_{7-\delta}$ (X= Ag, Al) samples. An increase in $J_C$ has been observed with Ag doping. Because of inaccessibility of higher magnetic fields in our lab, we could not measure $J_C$ of the $YBa_2Cu_{3-x}Ag_xO_{7-\delta}$ samples with x=0.15, 0.3; this implies that these samples had higher $J_C$ values. Furthermore, $J_C$ increases by increasing Al doping in the samples with x=0.01, 0.02, and then decreases. According to the Zeldov model for pinning energy U(J) [37], we obtained $J_C$(Al-doped)/$J_C$(Pure) for $YBa_2Cu_{3-x}Al_xO_{7-\delta}$ (x=0-0.045) at T=77K and H=9kOe. The results are presented in Table 3. It is seen that $J_C$ increases with the Al doping in the samples with x=0.01, 0.02, and then decreases. In the samples with x=0.03, 0.045, diminish of $J_C$ may be due to the undesired secondary phase $Y_2BaCuO_5$ in these samples.

In the normal state, we have investigated temperature dependence of the Hall charge carrier density − see Figs. 6a and 6b. We have observed that the density increases as Ag doping increases. In fact, substitution of $Ag^+$ for Cu (1) increases the Hall charge carrier density, especially in our optimum sample x=0.15. Moreover, the density is larger for x=0.3 in comparison to x=0.2. In the Al-doped samples, the density follows a downward trend by Al doping. This implies that substitution of $Al^{3+}$ for the Cu (1) localizes holes at the Cu-O chains.

In the vortex state, we have studied the anomalous sign reversal − see Figs. 7-9. To provide a better contrast in Fig. 7, in the Ag-doped samples the temperature dependence of the $R_H$ is represented separately at 4.61kOe. It is seen that all samples show a dip once at this magnetic field. Furthermore, as doping increases up to x=0.15, the magnitude of this dip increases, and for excess amount of Ag it decreases (although it is still larger than its value for pure samples). Figure 8 represents the T-dependence of $R_H$ in the Al-doped samples at 4.61kOe in the vortex state. One can see a single sign reversal in all samples, and magnitude of the dip has a downward trend up to the doping value x=0.02, and it is almost constant for higher concentrations (for which the samples contain small amount of Y211 impurity). Diminishing of the dip has also been reported in YBCO with doping of $BaZrO_3$ nanoparticles [38]. In contrast to our previous results for Ni- and Fe-doped YBCO samples [39], this dip has been observed at the same temperature for both Ag and Al doping. This implies that Ag and Al dopants do not cause any disturbance in superconductivity, nor do they cause any disorder in vortex motion.

According to our I-V result, Ag and Al doping (up to x=0.02) enhances $J_C$ and subsequently the pinning energy $U_J$. $J_C$, however, decreases for excess concentration of Al. In fact, it seems that these dopants and also small amount of Y211 impurity affect the pinning energy. Therefore, we can attribute these changes of the Hall dip to changes of acting forces on vortices, such as pinning force, which occurs by Ag doping [7, 40-42]. Figure 9a and 9b show the temperature dependence of the Hall coefficient in the Ag-doped sample with optimum amount x=0.15 and the Al-doped sample with x=0.02, which does not have any impurity, at H=2.52, 4.61, and 6.27kOe. The Hall dip is observed once at all three magnetic fields, and it is seen at the same temperature at various magnetic fields in contrast to our result for Ni- and Fe- doped samples [39]. Moreover, the magnitude of this dip decreases inversely with the value of magnetic field. Similar behavior has already been observed in YBCO, BSCCO [42] and $GdBa_{2-x}Nd_xCu_3O_{7-\delta}$ [43].

## 4. Conclusion and outlook

In summary, we have studied the Hall effect, anomalies associated to it, and transport properties of Y123 superconductors due to electron and hole doping (through Al and Ag substitution, respectively).

Specifically, we have measured electrical resistivity, Hall anomaly of Ag- and Al-doped YBCO samples at various magnetic fields. We have observed that the critical temperature $T_C$ decreases with Al doping, whereas for Ag doping $T_C$ increases so that it obtains its maximum for x=0.15. Similarly, the critical current density $J_C$ exhibited an increase for the Ag/Al doping up to x=0.15/0.02; while excessive concentrations decreased $J_C$. In addition, temperature dependence of the Hall charge carrier density showed an upward trend with Ag doping and a downward trend with Al doping. We also observed a single Hall sign reversal for all samples at H=4.61kOe. At this field value, the size of the Hall dip increases with both Ag doping and Al doping up to the values x=0.02 and 0.15, respectively; while the dip shrinks for larger concentrations. Moreover, the size of the Hall dip corresponding to the aforementioned optimal x-values decreases with magnetic field for the Ag-doped case; while for the Al-doped case it behaves conversely.

In a broader perspective, detailed studies as ours not only are of fundamental importance for understanding behavior of superconducting materials, but also they might provide some hints for applications in other areas. For instance, our capability of dealing with vortex dynamics controllably may suggest applications of appropriately-doped superconductors in devising electronic gates. In a different context, in systems with fractional quantum Hall effect, manipulation of vortices ("entangling" them, for example) is the main bottleneck for the possibility of using such materials in emergent fields such as quantum computation. We thus hope that our study can stimulate similar attempts to utilize manipulated superconducting materials for wider applications in, e.g., electronics.

## 5. Acknowledgments

Partial financial support by Alzahra University's vice president for research is gratefully acknowledged.

Figure captions:

Figuer 1. XRD patterns of a) $YBa_2Cu_{3-x}Ag_xO_{7-\delta}$ (x= 0-0.3) samples, and b) $YBa_2Cu_{3-x}Ag_xO_{7-\delta}$ (x= 0-0.045) samples.

Figure 2. SEM micrographs of $YBa_2Cu_{3-x}Ag_xO_{7-\delta}$ samples with different doping concentrations.

Figure 3. SEM micrographs of $YBa_2Cu_{3-x}Al_xO_{7-\delta}$ samples with different doping concentrations.

Figure 4. Temperature dependence of a) the electrical resistivity for the Ag-doped samples and $T_C$ vs. x (Ag doping), and b) the electrical resistivity for the Al-doped samples and $T_C$ vs. x (Al doping).

Figure 5. E-J curves for a) the Ag-doped samples, and b) the Al-doped samples at T=77K and H=9kOe.

Figure 6. Temperature dependence of a) the charge carrier density for the Ag-doped samples, and b) the charge carrier density for the Al-doped samples at H=4.61kOe.

Figure 7. Temperature dependence of the Hall coefficient of the $YBa_2Cu_{3-x}Ag_xO_{7-\delta}$ (x=0-0.3) samples at H=4.61kOe in the vortex state.

Figure 8. Temperature dependence of the Hall coefficient of the $YBa_2Cu_{3-x}Al_xO_{7-\delta}$ (x=0-0.045) samples at H=4.61kOe in the vortex state.

Figure 9. Temperature dependence of a) the Hall coefficient for Ag-doped sample x=0.15, and b) the Hall coefficient for Al-doped sample x=0.02 at various magnetic fields in the vortex state.

Table captions:

Table 1. Structural parameters and occupancy of $YBa_2Cu_{3-x}Ag_xO_{7-\delta}$ with (x=0-0.3).

Table 2. Structural parameters and occupancy of $YBa_2Cu_{3-x}Al_xO_{7-\delta}$ with (x=0-0.045).

Table 3. $J_C$ (Al-doped)/$J_C$ (Pure) for $YBa_2Cu_{3-x}Al_xO_{7-\delta}$ (x=0-0.045) in T=77K and H=9kOe.

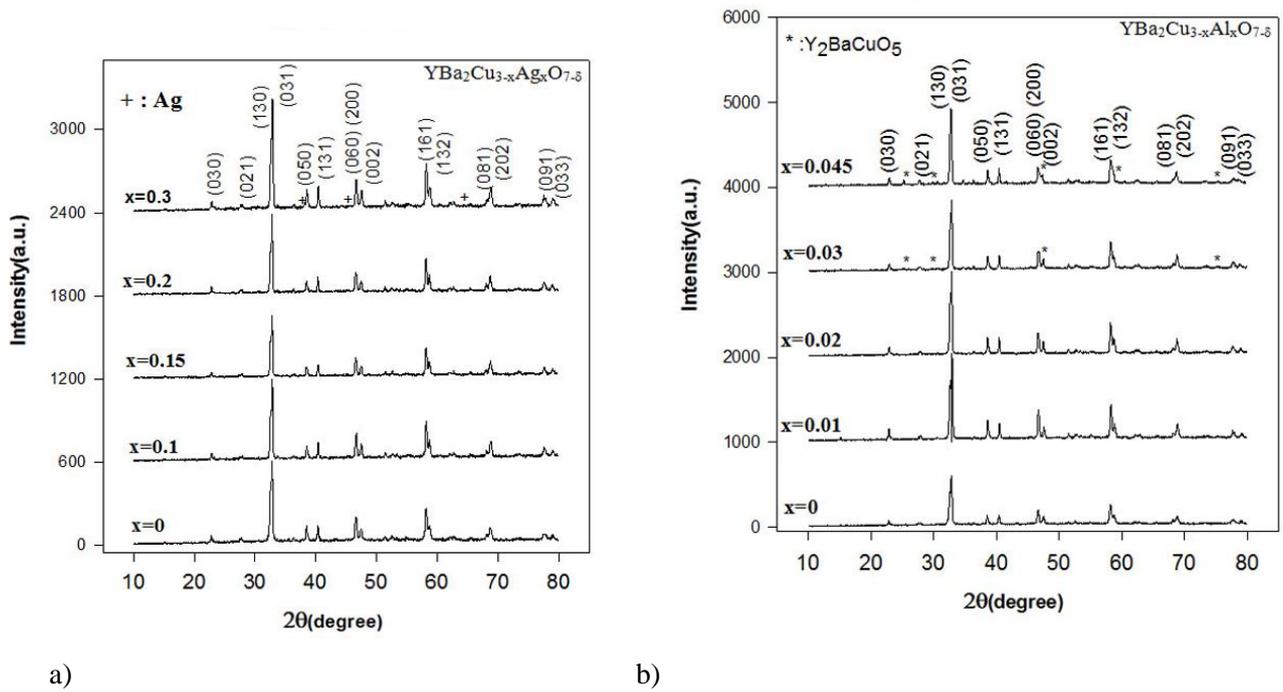

a)                                      b)

Fig. 1: XRD patterns of a) YBa$_2$Cu$_{3-x}$Ag$_x$O$_{7-\delta}$ (x= 0-0.3) samples, and b) YBa$_2$Cu$_{3-x}$Ag$_x$O$_{7-\delta}$ (x= 0-0.045) samples.

Substitution effects of Ag/Cu and Al/Cu on Y123 samples: Hall anomaly and transport properties

M. Nazarzadehmoafi[1], V. Daadmehr[*1], A. T. Rezakhani[2], S. Falahati[1], F. Saeb[1], and S. Barekat Rezaei[1]

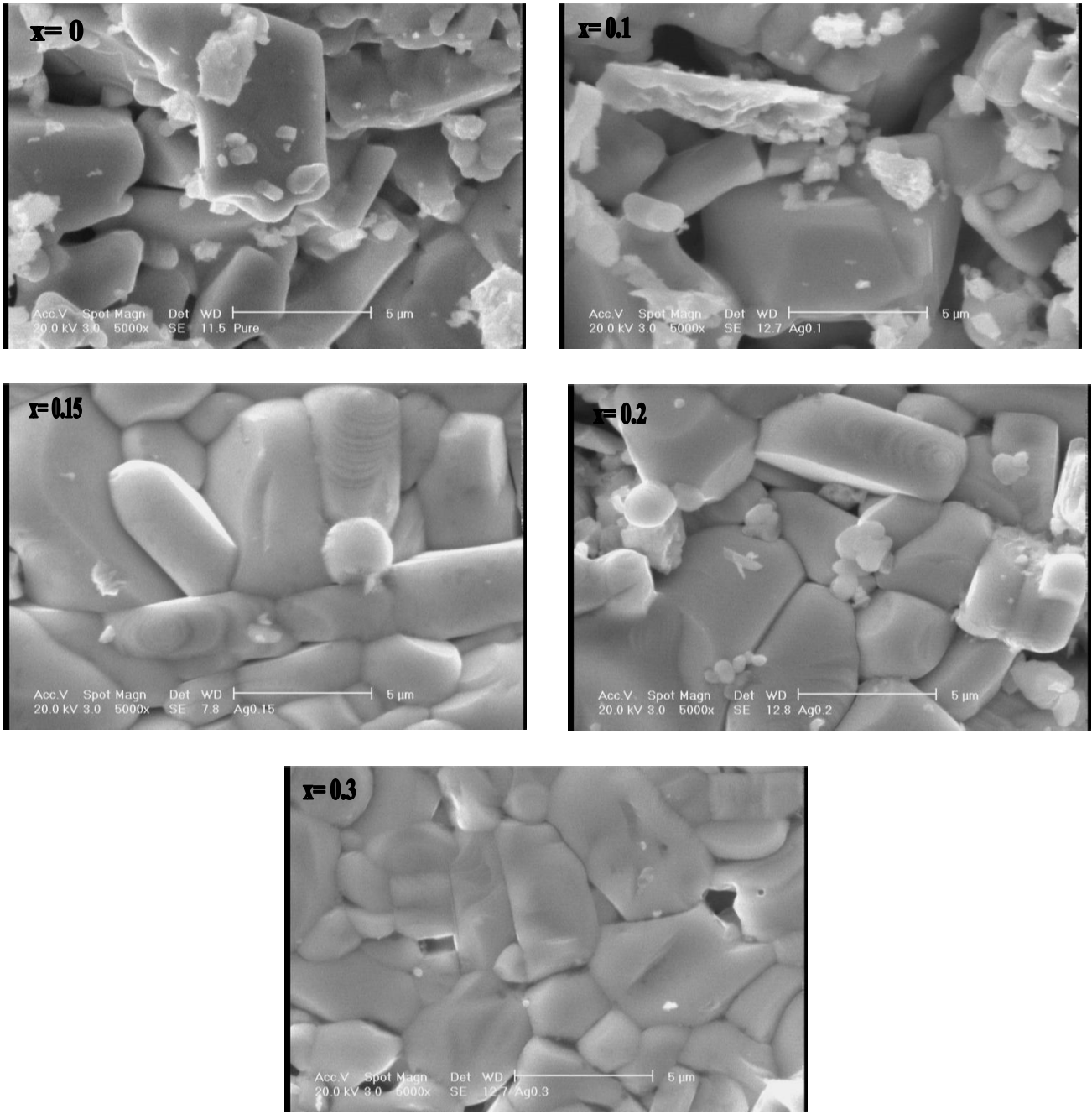

Fig. 2: SEM micrographs of YBa$_2$Cu$_{3-x}$Ag$_x$O$_{7-\delta}$ samples with different doping concentrations.

Substitution effects of Ag/Cu and Al/Cu on Y123 samples: Hall anomaly and transport properties

M. Nazarzadehmoafi[1], V. Daadmehr[*1], A. T. Rezakhani[2], S. Falahati[1], F. Saeb[1], and S. Barekat Rezaei[1]

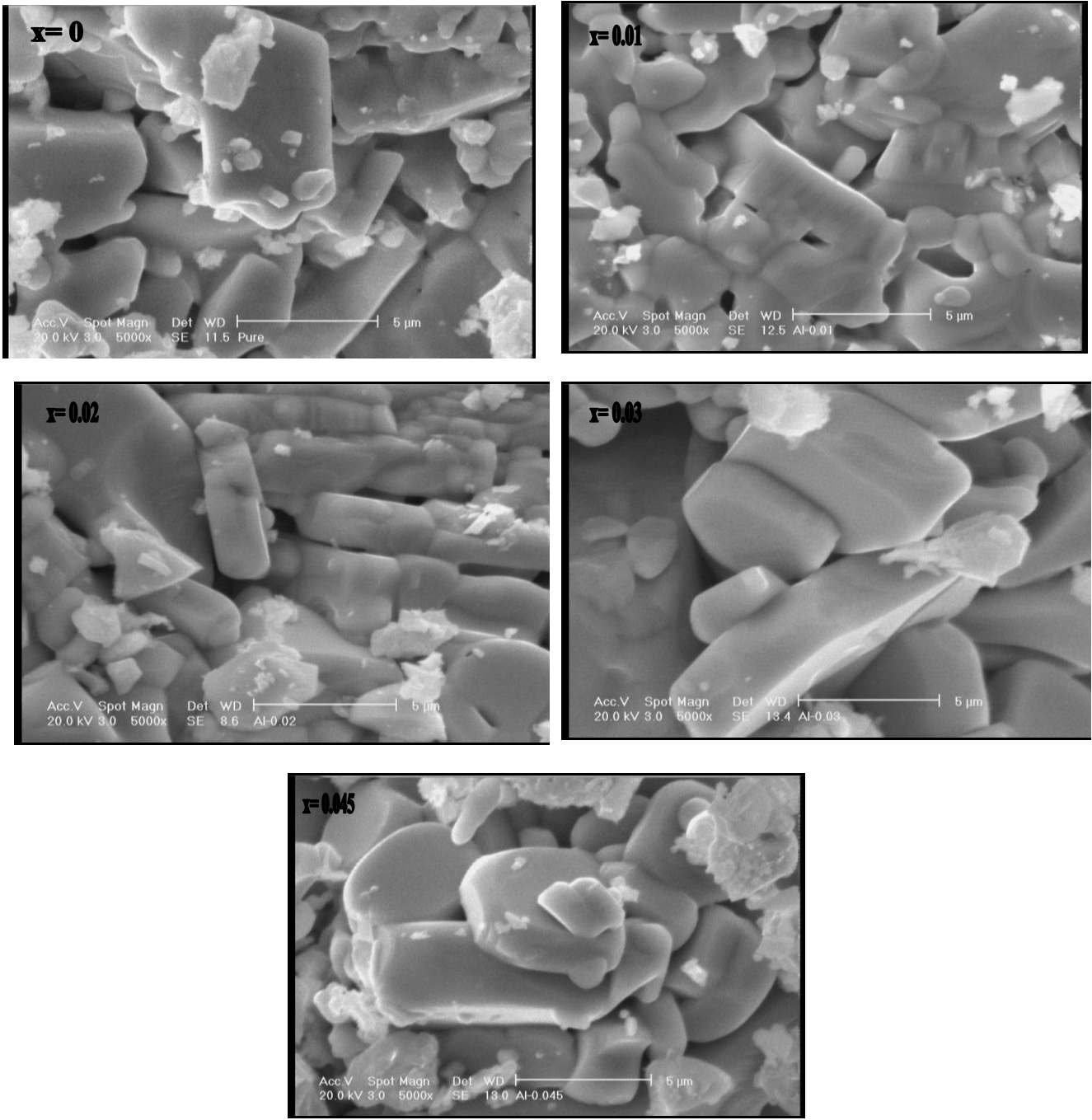

Fig. 3: SEM micrographs of $YBa_2Cu_{3-x}Al_xO_{7-\delta}$ samples with different doping concentrations.

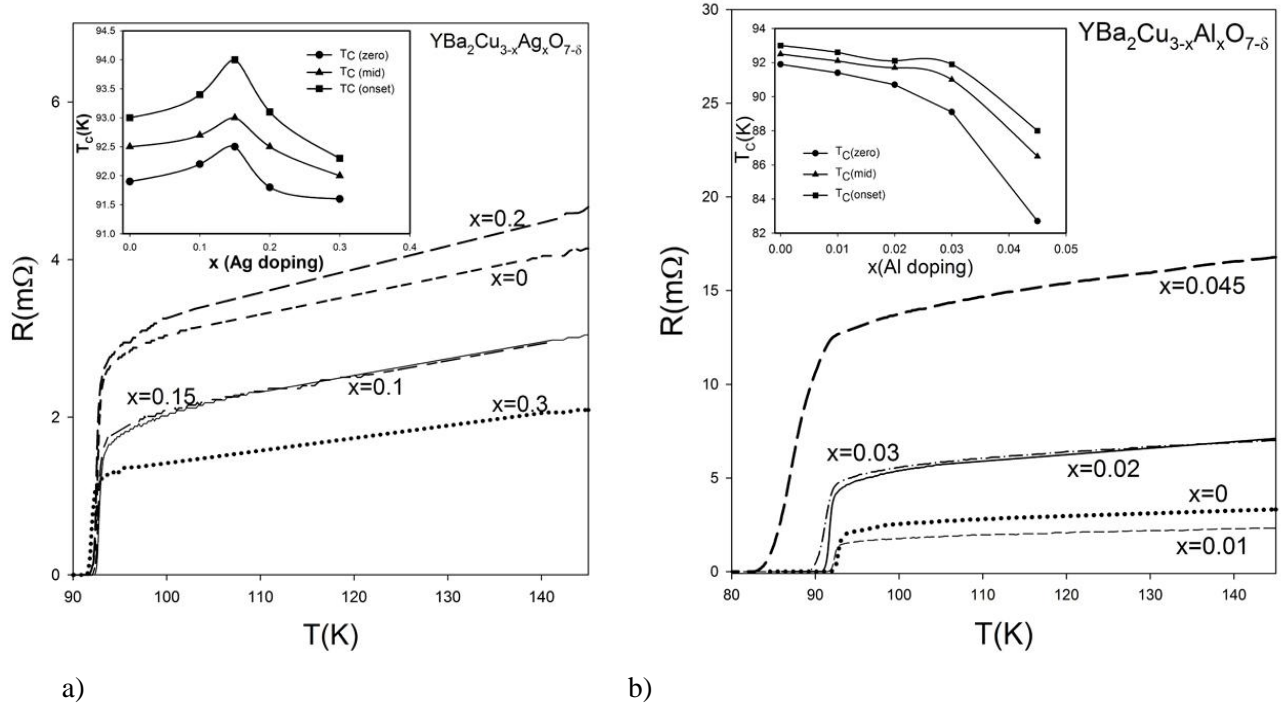

Fig.4: Temperature dependence of a) the electrical resistivity for the Ag-doped samples and $T_C$ vs. x (Ag doping), and b) the electrical resistivity for the Al-doped samples and $T_C$ vs. x (Al doping).

Substitution effects of Ag/Cu and Al/Cu on Y123 samples: Hall anomaly and transport properties

M. Nazarzadehmoafi[1], V. Daadmehr[*1], A. T. Rezakhani[2], S. Falahati[1], F. Saeb[1], and S. Barekat Rezaei[1]

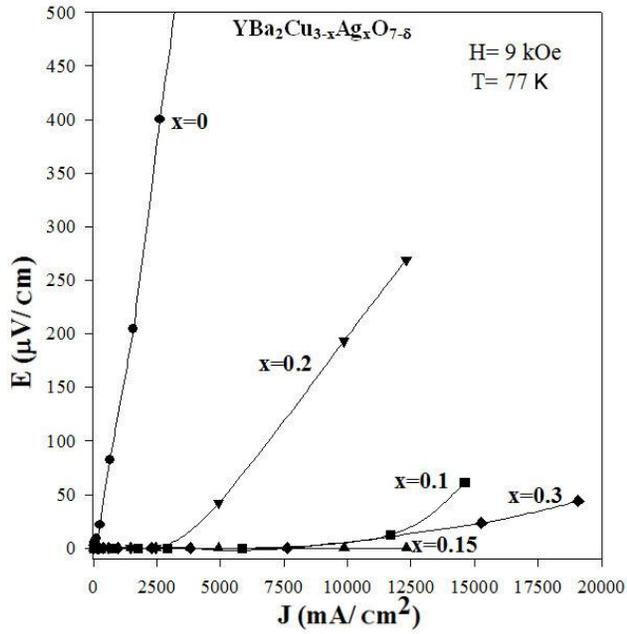 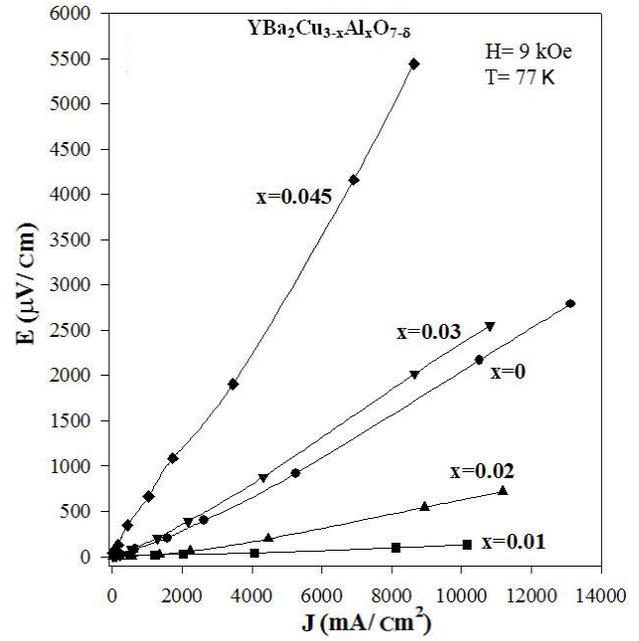

a) b)

Fig.5: E-J curves for a) the Ag-doped samples, and b) the Al-doped samples at T=77K and H=9kOe.

Substitution effects of Ag/Cu and Al/Cu on Y123 samples: Hall anomaly and transport properties

M. Nazarzadehmoafi[1], V. Daadmehr[*1], A. T. Rezakhani[2], S. Falahati[1], F. Saeb[1], and S. Barekat Rezaei[1]

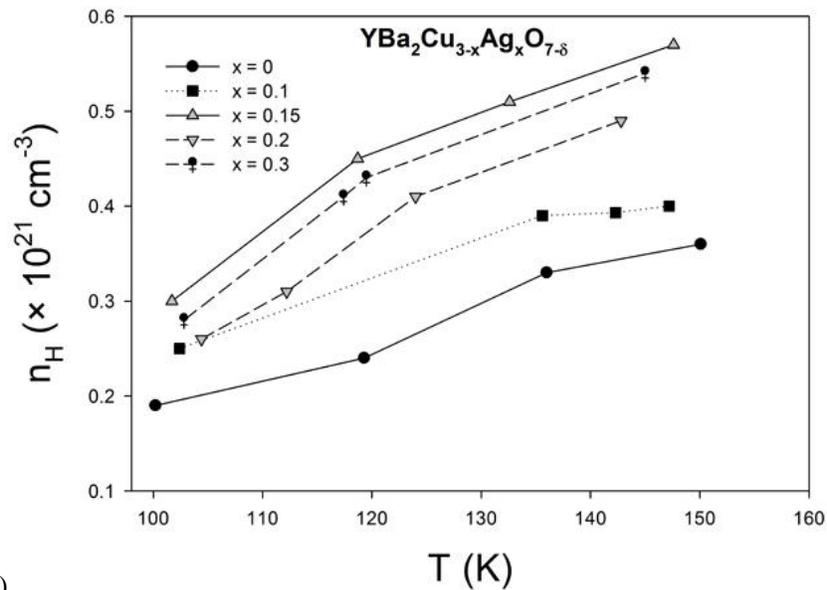

a)

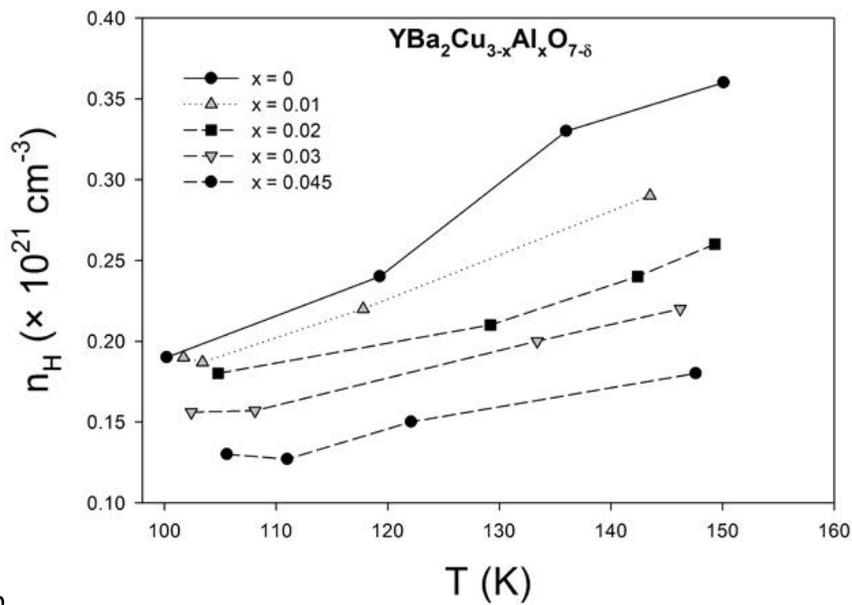

b

Fig. 6: Temperature dependence of a) the charge carrier density for the Ag-doped samples, and b) the charge carrier density for the Al-doped samples at H=4.61kOe.

Substitution effects of Ag/Cu and Al/Cu on Y123 samples: Hall anomaly and transport properties

M. Nazarzadehmoafi[1], V. Daadmehr[*1], A. T. Rezakhani[2], S. Falahati[1], F. Saeb[1], and S. Barekat Rezaei[1]

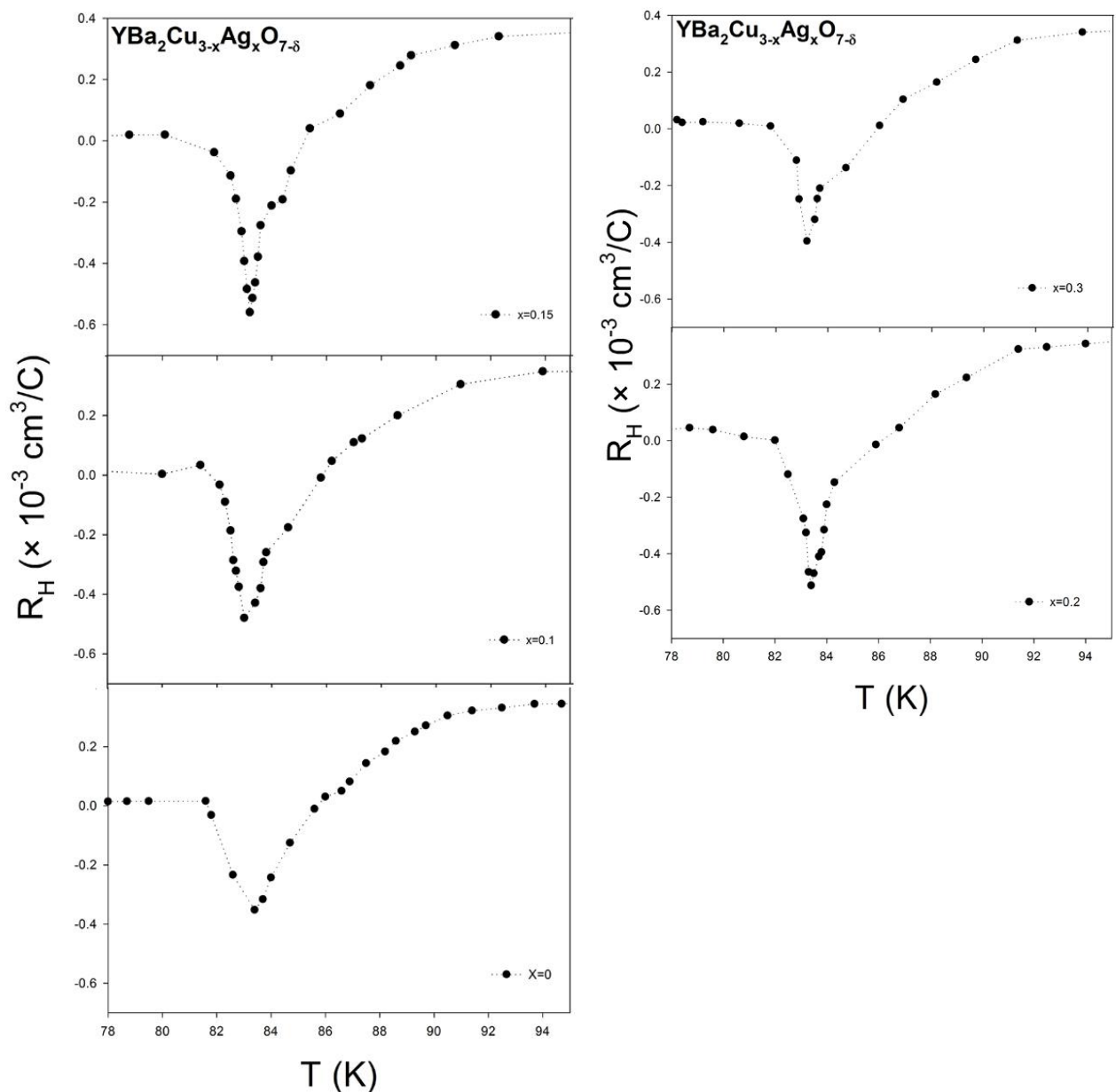

Fig. 7: Temperature dependence of the Hall coefficient of the $YBa_2Cu_{3-x}Ag_xO_{7-\delta}$ (x=0-0.3) samples at H=4.61kOe in the vortex state.

Substitution effects of Ag/Cu and Al/Cu on Y123 samples: Hall anomaly and transport properties

M. Nazarzadehmoafi[1], V. Daadmehr[*,1], A. T. Rezakhani[2], S. Falahati[1], F. Saeb[1], and S. Barekat Rezaei[1]

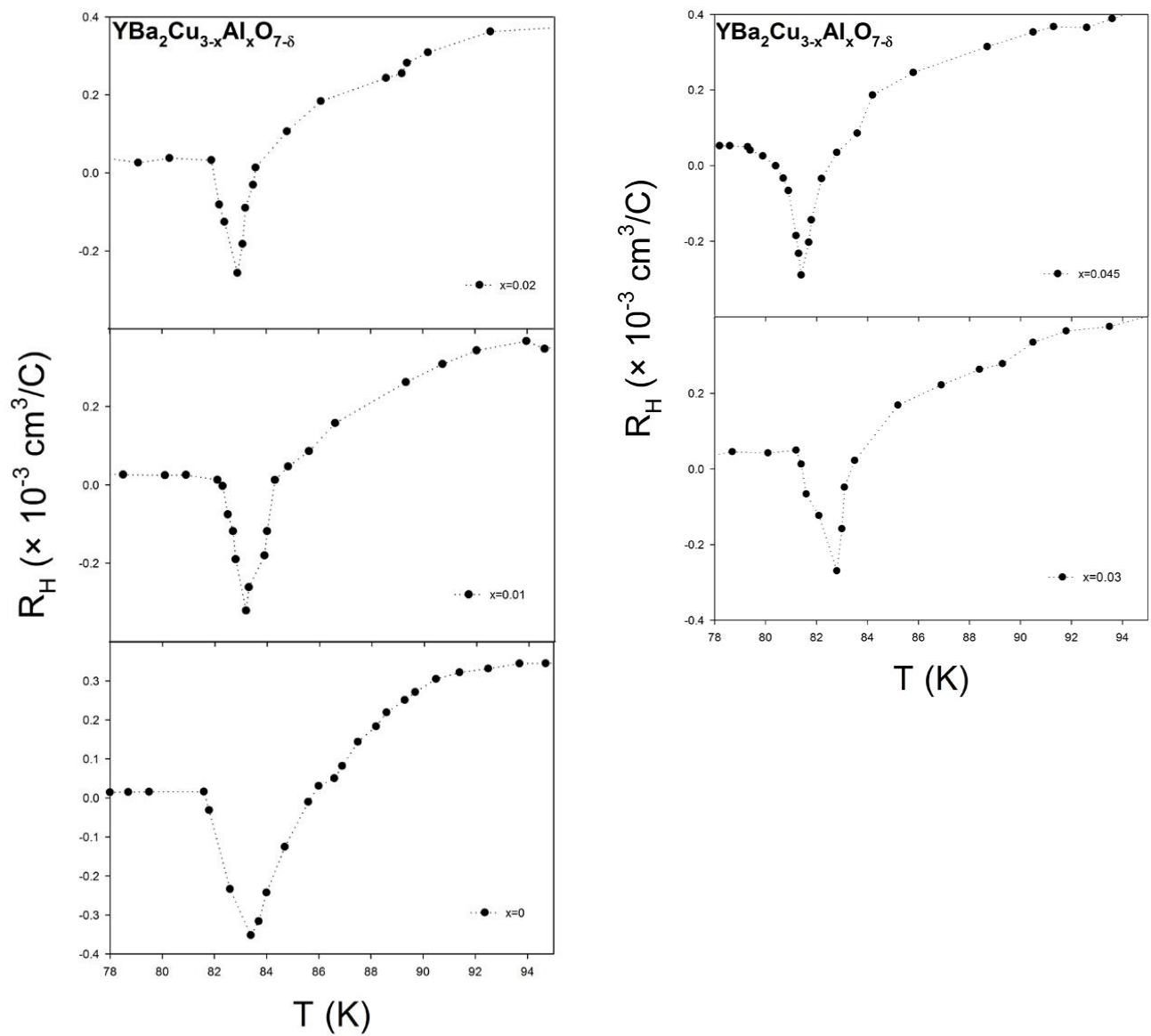

Fig. 8: Temperature dependence of the Hall coefficient of the $YBa_2Cu_{3-x}Al_xO_{7-\delta}$ (x=0-0.045) samples at H=4.61kOe in the vortex state.

Substitution effects of Ag/Cu and Al/Cu on Y123 samples: Hall anomaly and transport properties

M. Nazarzadehmoafi[1], V. Daadmehr[*1], A. T. Rezakhani[2], S. Falahati[1], F. Saeb[1], and S. Barekat Rezaei[1]

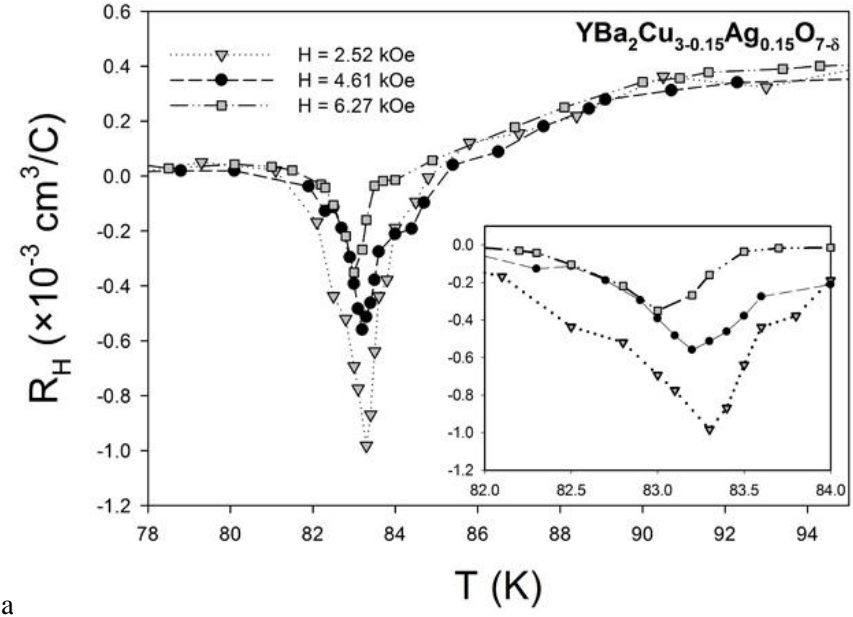

a

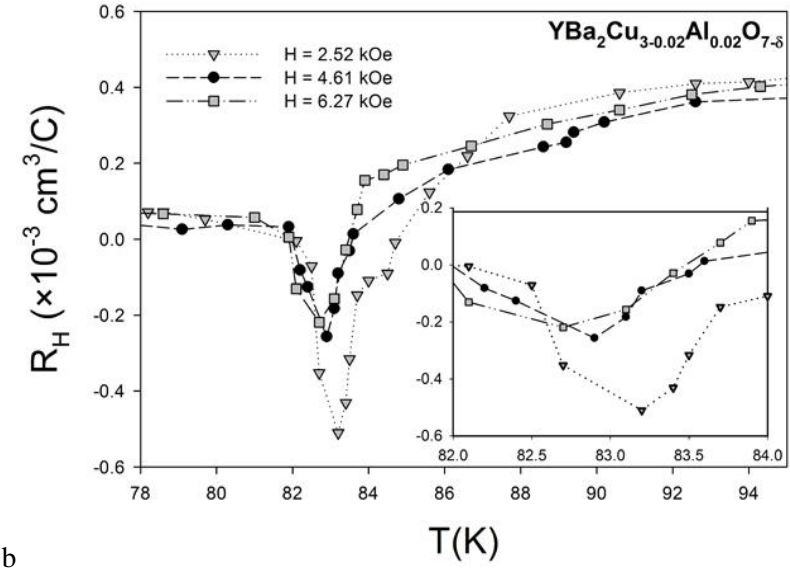

b

Fig. 9: Temperature dependence of a) the Hall coefficient for Ag-doped sample x=0.15, and b) the Hall coefficient for Al-doped sample x=0.02 at various magnetic fields in the vortex state.

Substitution effects of Ag/Cu and Al/Cu on Y123 samples: Hall anomaly and transport properties

M. Nazarzadehmoafi[1], V. Daadmehr[*1], A. T. Rezakhani[2], S. Falahati[1], F. Saeb[1], and S. Barekat Rezaei[1]

Table 1. Structural parameters and occupancy of $YBa_2Cu_{3-x}Ag_xO_{7-\delta}$ with (x=0-0.3).

| $YBa_2Cu_{3-x}Ag_xO_7$ | x=0 | x=0.1 | x=0.15 | x=0.2 | x=0.3 |
|---|---|---|---|---|---|
| a(Å) | 3.8202963 | 3.8203027 | 3.8203108 | 3.8203201 | 3.8202949 |
| b(Å) | 3.8854918 | 3.8854802 | 3.885493 | 3.8855014 | 3.8854713 |
| c(Å) | 11.6835375 | 11.683507 | 11.683557 | 11.683595 | 11.683476 |
| V(Å$^3$) | 173.4273 | 173.4266 | 173.4283 | 173.4296 | 173.4254 |
| orthorombicity | 0.8461 | 0.8458 | 0.8459 | 0.8459 | 0.8458 |
| Y (z position) | 0.4995 | 0.4995018 | 0.49950004 | 0.4995 | 0.49968368 |
| Ba (z position) | 0.18397921 | 0.1839669 | 0.18396258 | 0.18397489 | 0.18395264 |
| Cu(1) | | | | | |
| occupancy | 1.00 | 0.8986215 | 0.848373 | 0.7977579 | 0.6994978 |
| z position | -0.000052 | -0.0001369 | -0.0000935 | -0.000059 | -0.0000806 |
| Cu(2) | | | | | |
| occupancy | 1 | 0.99937505 | 0.99911636 | 0.9987758 | 0.9999272 |
| z position | 0.3551134 | 0.3550506 | 0.3550452 | 0.35506198 | 0.3550262 |
| Ag | | | | | |
| occupancy | - | 0.09921873 | 0.1490817 | 0.19872798 | 0.2997142 |
| z position | - | 0.00302 | 0.0000633 | -0.0000328 | 0.0000126 |
| O(1) | | | | | |
| occupancy | 0.91666967 | 0.9133287 | 0.9124455 | 0.91390324 | 0.9999272 |
| z position | -0.0000466 | -0.000119 | 0.0000002 | -0.000152 | 0.0001299 |
| O(4) (z position) | 0.15882613 | 0.1584899 | 0.15849802 | 0.15854175 | 0.15849113 |

Substitution effects of Ag/Cu and Al/Cu on Y123 samples: Hall anomaly and transport properties

M. Nazarzadehmoafi[1], V. Daadmehr[*1], A. T. Rezakhani[2], S. Falahati[1], F. Saeb[1], and S. Barekat Rezaei[1]

Table 2. Structural parameters and occupancy of $YBa_2Cu_{3-x}Al_xO_{7-\delta}$ with (x=0-0.045).

| $YBa_2Cu_{3-x}Al_xO_7$ | x=0 | x=0.01 | x=0.02 | x=0.03 | x=0.045 |
|---|---|---|---|---|---|
| a(Å) | 3.8202963 | 3.8203638 | 3.8203068 | 3.8203049 | 3.835369 |
| b(Å) | 3.8854918 | 3.8856194 | 3.8854742 | 3.8854673 | 3.881315 |
| c(Å) | 11.6835375 | 11.683748 | 11.683487 | 11.68347 | 11.696847 |
| V(Å$^3$) | 173.4273 | 173.4392 | 173.4262 | 173.4256 | 174.1225 |
| orthorombicity | 0.8461 | 0.8468 | 0.8457 | 0.8456 | 0.5954 |
| Y (z position) | 0.4995 | 0.50048965 | 0.4995 | 0.4995 | 0.4498944 |
| Ba (z position) | 0.18397921 | 0.1841467 | 0.18396032 | 0.18395934 | 0.19333248 |
| Cu(1) | | | | | |
| occupancy | 1.00 | 0.98376703 | 0.9797354 | 0.96961737 | 0.80789185 |
| z position | -0.000052 | -0.000187 | -0.000109 | -0.0000009 | -0.0451525 |
| Cu(2) | | | | | |
| occupancy | 1 | 0.9966223 | 1.0003636 | 1.0001091 | 1.000 |
| z position | 0.3551134 | 0.355514 | 0.35504037 | 0.35504282 | 0.35036322 |
| Al | | | | | |
| occupancy | - | 0.007851183 | 0.019531965 | 0.029051231 | 0.08341494 |
| z position | - | -0.03066187 | -0.00950515 | -0.000065 | -0.05034 |
| O(1) | | | | | |
| occupancy | 0.91666967 | 0.8868087 | 0.9117832 | 0.9125029 | 2.8149834 |
| z position | -0.0000466 | -0.000498 | 0.0000029 | 0.0000008 | -0.024368 |
| O(4) (z position) | 0.15882613 | 0.16034253 | 0.15842703 | 0.15841465 | 0.15282746 |

Substitution effects of Ag/Cu and Al/Cu on Y123 samples: Hall anomaly and transport properties

M. Nazarzadehmoafi[1], V. Daadmehr[*1], A. T. Rezakhani[2], S. Falahati[1], F. Saeb[1], and S. Barekat Rezaei[1]

Table 3. $J_C$ (Al-doped)/$J_C$ (Pure) for $YBa_2Cu_{3-x}Al_xO_{7-\delta}$ (x=0-0.045) in T=77K and H=9kOe.

| sample | $J_C$(Al-doped)/$J_C$(Pure) |
|---|---|
| $YBa_2Cu_3O_{7-\delta}$ | 1 |
| $YBa_2Cu_{2.99}Al_{0.01}O_{7-\delta}$ | 18.4260 |
| $YBa_2Cu_{2.98}Al_{0.02}O_{7-\delta}$ | 10.7012 |
| $YBa_2Cu_{2.97}Al_{0.03}O_{7-\delta}$ | 0.8613 |
| $YBa_2Cu_{2.955}Al_{0.045}O_{7-\delta}$ | 0.2371 |

Substitution effects of Ag/Cu and Al/Cu on Y123 samples: Hall anomaly and transport properties

M. Nazarzadehmoafi[1], V. Daadmehr[*,1], A. T. Rezakhani[2], S. Falahati[1], F. Saeb[1], and S. Barekat Rezaei[1]